\documentstyle[12pt,epsf]{article}
\input epsf
\textheight 9in  \topmargin -.4in   \def\baselinestretch{1.4}
\def\Eq{\begin{equation}}	\def\End{\end{equation}}
\def\Eqa{\begin{eqnarray}}	\def\Enda{\end{eqnarray}}
\def\Endl#1{\label{#1} \End}	\def\Endla#1{\label{#1} \Enda}
	
		\def\to{\!\rightarrow\!}
\def\etal{{\it et.al.}}

\def\frac#1#2{{\textstyle {#1 \over #2}}}

\def\dspace#1{\renewcommand{\baselinestretch}{#1} \large\normalsize}
\def\dspace#1{\relax} 
\def\lqcd{ \Lambda_{\rm QCD} }
\def\gev{ {\rm GeV} }
\def\Sum{ \sum_{X \ne H} \!\! ^\prime\ }
\def\lesim{ \raisebox{-0.5ex}{$\stackrel{\textstyle <}{\sim}$} }

\def\eps{\epsilon}
\def\btorhop{$\overline B^0 \to \rho^+ l^- \overline \nu$}
\def\btorho{$\overline B \to \rho l \overline \nu$}

\def\btopi{$\overline B \to \pi l \overline \nu$}

\def\Lam{\bar \Lambda}

\def\OMIT#1{}

\def\etal{{\it et al.\/}}

\def\CO{{\cal O}}
\def\frac#1#2{{#1\over#2}}

\def\np#1#2#3{\NP{\bf B#1} (#2) #3}
\def\pl#1#2#3{\PL B {\bf #1} (#2) #3}

\def\pr#1#2#3{\PR{\bf #1} (#2) #3}

\begin{document}
\begin{titlepage}
\rightline{CMU-HEP-97-12}
\rightline{hep-ph/yymmxxx}
\rightline{November 1997}
\vskip.5in
\begin{center}

{\Large \bf
The Phenomenology of Inclusive Heavy-to-Light Sum Rules
}
\vskip.2in
{
  {\bf C. Glenn Boyd}\footnote{\tt boyd@fermi.phys.cmu.edu}\\
\vskip.2cm
  {\bf  I.Z. Rothstein}\footnote{\tt ira@cmuhep2.phys.cmu.edu}\\
  \vskip.2cm {\it Department of Physics}\\
  \vskip.2cm {\it Carnegie Mellon University}\\
  \vskip.2cm {\it Pittsburgh, Pa 15213-3890}}
\vskip.2cm
\end{center}

\vskip.5in
\begin{abstract}
\dspace{1.4}

By calculating the $O(\alpha_s)$ corrections to inclusive heavy-to-light
sum rules we find model independent upper and lower bounds on form 
factors for \btopi\ and \btorho. 
We use the bounds to rule out model predictions.
Some models violate the bounds only for certain ranges of sum rule
input parameters $\Lam \simeq m_B-m_b$ and $\lambda_1$, or for certain
choices of model parameters, while others obey or violate
the bounds irrespective of the inputs. We 
discuss the reliability and convergence of the bounds, point
out their utility for extracting $V_{ub}$, and 
derive from them a new form factor scaling relation.

\par
\end{abstract}
\end{titlepage}
\dspace{2.0}
\setcounter{footnote}{0}

\section{Introduction }

A precise extraction of the Cabibo-Kobayashi-Maskawa parameters 
is predicated on our ability
to control the theoretical uncertainties involved in the predictions of
heavy hadron decays. As such, it is imperative that we perform
our calculations within a systematic expansion. Heavy quark effective
theory\cite{hqet} leads to one such approximation scheme. It leads to 
predictions for the 
inclusive heavy hadron decay rates and spectra\cite{chay,incl}\ as an expansion in
$\alpha_s$ and $\Lambda_{QCD}/m_b$, as well as predictions for heavy to heavy
exclusive rates at zero recoil. Thus, with our present theoretical methods
we are capable of extracting $V_{cb}$ at the level of
a few percent. However, prospects for measuring $V_{ub}$ are much 
bleaker.  The standard method of extraction based on the lepton endpoint 
spectrum will likely be plagued by large theoretical uncertainties\cite{endpt},
while an extraction from the hadronic spectrum\cite{hadspec} 
will  suffer from the difficulty
of the measurement itself. Given that exclusive modes have already been measured,
it would be nice if we could get a handle on the exclusive 
form factors themselves. Unfortunately, our ability to calculate in QCD is limited
to inclusive quantities due to the fact that quark-hadron duality
is in itself an inclusive concept. If we attempt to   completely
eliminate the
excited states from our calculation we begin to probe the dynamics of
confinement, of which we are effectively ignorant. 
However, if we do not eliminate 
the excited states, but merely note that their contribution is positive
definite, then we can derive bounds on the form factors instead of predictions.

This idea was originally formulated by Bjorken to derive
upper bounds\cite{Bj} on form factors as a function 
of momentum transfer $q^2$, including a well-known bound on the slope of the 
Isgur-Wise function at zero recoil. In addition, Voloshin\cite{volo} was 
able to find a lower bound at zero recoil.  A few years later,
 Bigi \etal\cite{BSUV2}  
showed that these bounds can be derived from sum rules
similar to those developed for deep inelastic scattering, and are
in fact the leading order term in a  systematic expansion
in $\alpha_s$ and $\Lambda_{QCD}/m_b$. These sum rules were then used 
to refine bounds on heavy-to-heavy matrix elements at or near zero 
recoil\cite{BSUV2,BSUV,GLRW,GKLW,BGM,LS} and were applied as well to
heavy-to-light form factors far from zero recoil \cite{BR1}. In the
latter reference, momentum-dependent lower bounds were also developed.

These bounds are useful for several reasons. They may be used  
to  fine-tune and discriminate among  models 
used extensively for CKM extractions and  Monte-Carlo simulations of
backgrounds. Given the measured $q^2$ dependence of a form factor,
the upper and lower bounds may be used to extract a value 
for $V_{ub}$ without relying on models.  Since model
predictions for \btopi\ and \btorho\ decay rates vary by more than a 
factor of four, it is reasonable to ask whether comparing the total rate
to the integrated bounds may yield a more accurate extraction of $V_{ub}$.
We will see that the \btopi\ sum rule can place a more restrictive
lower bound on $V_{ub}$ than can be obtained by the union of models.
An upper bound may be derived as well using recent work\cite{disp,bglLam}
describing the shape of form factors to handle the 
problematic zero recoil region.

In this letter we continue the work begun in ref. \cite{BR1}
by including the effect of radiative corrections on the form factor bounds.
These effects are of leading order for the lower bounds, and in one
case  dominate the upper bound.  Due to the length
of the calculations we only present partial results here and leave
the more technical aspects for a subsequent, longer publication, where we will
present the full one loop radiative corrections to the hadronic tensor $T_{\mu \nu}$
defined by 
\Eqa
T^{\mu\nu}(v\cdot q, q^2) &=& \frac{-i}{2 M_B}\int d^4 x\ e^{-i q \cdot x} \langle B(v)| 
T\left(
J^{\mu\,\dagger}\left(x\right) J^\nu\left(0\right)\right)
\left|{B(v)}\right\rangle\nonumber\\ &\equiv&-g^{\mu\nu} T_1 +
v^{\mu}v^{\nu} T_2 - i \epsilon^{\mu\nu\alpha\beta} v_\alpha q_\beta
T_3 + q^\mu q^\nu T_4\nonumber\\ && + \left(q^\mu v^\nu + q^\nu
v^\mu\right) T_5\ ,
\Endla{tdef}
where $J$ is a flavor-changing current, $v^\mu$ is the $B$ meson
four-velocity, and $q^\mu$ is the dilepton four-momentum.
These corrections are useful in their own right since 
they can be used to calculate physical rates 
with arbitrary cuts, via a simple numerical integration.

Let us briefly review the derivation of the bounds. We follow the notation
used in \cite{BR1} and refer the reader to this reference for details.
The idea is to equate the inclusive rate, which is calculable
within perturbative QCD using an operator product expansion, 
with the sum over exclusive states. 
The sum over exclusive states is then truncated, 
and the equality is changed to an inequality.
More explicitly, equating partonic and hadronic sums over intermediate
states in Eq.~(\ref{tdef}), we have
\Eqa \label{masterSR}
&&{ | \langle H| a\cdot J | B \rangle |^2 \over 4 M_B E_H \eps }
+ \Sum { | \langle X| a\cdot J | B \rangle |^2 \over 2 M_B 
(\eps + E_H -E_X)} \nonumber \\
&&- \sum_X (2\pi)^3 \delta^{(3)}(\vec p_X
          - \vec q) { | \langle B| a\cdot J | X \rangle |^2 \over 2 
M_B (\eps + E_H +E_X -2 M_B)} 
= a^{\mu*} T^{OPE}_{\mu\nu} a^\nu .
\Endla{aTa}
Here $a_\mu$ is an arbitrary four-vector chosen to pick out the form
factor of interest for the $B\rightarrow H$ transition (eventually we will
take $H$ to be a pion or rho), $\epsilon$ is defined 
as $\epsilon = M_B-E_H-v\cdot q$, and  $E_H=\sqrt{M_H^2+ \vec q^2}$.
The first two terms represent the local cut in the complex $v\cdot q$ plane
corresponding to the semi-leptonic decay process. The third term arises from 
the ``distant cut,'' corresponding to the pair production process.
The sum over states contains the usual phase space integration
$\int d^3p/ (2 E)$ for each particle, while $\Sigma_{X \ne H}'$ is shorthand for
\Eq
\Sum \equiv \sum_{X \ne H}\, (2\pi)^3 \delta^{(3)}
(\vec p_X
          + \vec q).\nonumber
\Endl{sigdef}
To justify the use of the OPE on the right hand side of Eq.~(\ref{aTa}), we 
perform a contour integral over $\epsilon$ with a weighting function as 
described in \cite{GKLW}. We choose the weighting function $W_\Delta(\eps)$ to be
$W=\theta (\Delta-\epsilon)$ \footnote{In ref. \cite{GKLW} it was shown 
that for $b \to c$  the results depend rather mildly on the precise 
form of the weight function.}. Delta has
the effect of cutting off the contribution from excited states, which in turn
improves the bound beyond what one would get trivially
from insisting that the one exclusive mode be less than the
inclusive rate. However, as discussed above it is not possible to choose
$\Delta$ to be too small, since as we eliminate the contributions from 
excited states we become more sensitive to hadronization effects, and
errors due to duality violation become important. Thus, we must choose
$\Delta$ small enough to maximize the utility of our bounds, 
yet large enough to preserve their validity, $\Lambda_{QCD}<<\Delta<m_b$,
where $\lqcd$ is a typical hadronic scale. Furthermore, the OPE
is an expansion in $\Lambda_{QCD}/E_H$, so we can expect it to converge
only when the hadronic three momentum $q_3 \equiv |\vec q|$ 
is sufficiently large, $q_3 >> \Lambda_{QCD}$.

By appropriate choice of  contour we may  eliminate the
contribution from the unphysical cut. The  upper bound may then be found
by noticing that the excited state contribution is positive definite,
leading to
\Eq
{ | \langle H| a\cdot J | B \rangle |^2 \over 4 M_B E_H }
\leq  \int d\eps\ W_\Delta(\eps)\ a^{\mu*} T^{OPE}_{\mu\nu} 
a^\nu .
\Endl{zmoment}
A lower bound can found by using the fact that
\Eq
(E_1 - E_H) \Sum { | \langle X| a\cdot J | B \rangle |^2 }
          W_\Delta(E_X -E_H)  \le
 \ \ \Sum {(E_X - E_H) | \langle X| a\cdot J | B \rangle |^2 }
          W_\Delta(E_X -E_H) \ ,
\Endl{voltrick}
where $E_1$ is the energy of the first excited state more massive than $H$. 
The contribution of multi-particle states with energies less 
than that of the first excited resonance has been neglected, as 
they are suppressed by both phase space and large-$N_c$ power counting.
We therefore have both upper and lower bounds,
\Eq \label{bnds}
\int d\eps\ W_\Delta(\eps)\ a^{\mu*} T^{OPE}_{\mu\nu} a^\nu \ge 
{| \langle H| a\cdot J | B \rangle |^2 \over 4 M_B E_H }\ge 
\int d\eps\ W_\Delta(\eps)\ a^{\mu*} T^{OPE}_{\mu\nu} a^\nu 
 \biggl[ 1 - { \eps\over E_1 - E_H} \biggr]  .
\Endl{generalbd}

Since Eq.~(\ref{masterSR}) forces $\epsilon = E_X -E_H$,
we see that the last term above may be interpreted
as the average excitation energy of the higher states
contributing to the sum rule, weighted by $W_\Delta(\eps) \  a^\dagger T a$.
As emphasized in \cite{BR1}, the excitation energy begins at ${\cal O}(\lqcd)$
and receives radiative corrections of order $\alpha_s \Delta$. Since
these two quantities are numerically comparable, the lower bounds
are not trustworthy without the radiative corrections, which we
now include. 
Furthermore,
given that our bounds are only valid when $q_3>>\lqcd$, we see
that the lower bound will only be useful if 
\Eq
{ (C_1 \lqcd +C_2 \alpha_s
\Delta)\ q_3 \over (M_1^2-M_H^2)} \lesim \ 1.
\End
Thus, we see again there is a trade off between the utility and validity 
of the lower bound, although we are helped in this case by
the fact that typical hadronic masses $\sim 1 \gev$ are numerically
larger than nonperturbative condensates such as $\Lam$. 
The range in $q_3$ for which
the lower bounds are both useful and reliable depends on the
nonperturbative corrections $C_1 \lqcd$ presented in \cite{BR1}\
and the radiative correction $C_2 \alpha_s \Delta$ presented here.
We will see that the usefulness
of the lower bound depends sensitively on the numerical value of
the heavy quark parameter $\bar{\Lambda}$.

The values of the parameters  $\bar{\Lambda}$ and $\lambda_1$
have been extracted from the shape of the lepton end-point spectrum
in inclusive semileptonic $B\rightarrow X l \bar{\nu_l}$ decay 
\cite{gremm,FLS}. The values found were $\bar{\Lambda}=0.39 \pm.11~\gev$ and
$\lambda_1=-0.19\pm 0.10~\gev^2$ (the uncertainty being the $1\sigma$ 
statistical error only). The errors are strongly correlated, and
therefore we will in general evaluate our bounds for three values of the pair 
$(\bar{\Lambda},\lambda_1)$: 
\Eqa
{\rm Set\ A}\ &\equiv&\ (\Lam = 0.28~\gev,\ \lambda_1=-0.09~\gev)
\nonumber\\
{\rm Set\ B}\ &\equiv&\ (\Lam=0.39~\gev,\  \lambda_1=-0.19~\gev)
\nonumber\\
{\rm Set\ C}\ &\equiv&\ (\Lam= 0.50~\gev,\ \lambda_1=-0.29~\gev)
\Enda
The value of
$\lambda_2 = 0.12\ \gev^2$ is determined from the $B-B^*$ hyperfine
splitting. Eventually, $\Lam$ may be extracted from measurements of
semileptonic $\Lambda_b \to \Lambda_c$ decays, because all form factors relevant
to this decay can be expressed in terms of $\Lam$ and one universal 
function\cite{GGW}, and this function is furthermore determined by 
a single ``shape parameter''\cite{bglLam}. Information on $\Lam$ and $\lambda_1$
may also be extracted from the $\bar B \to X_s \gamma$ spectrum.

\section{Heavy-to-Light Sum Rules}

Let us begin by considering the bounds on the $B \to \pi l \bar \nu$
form factor $f+$ defined by
\begin{equation}
\langle \pi(p^\prime)\mid \bar u\gamma^\mu b \mid \bar B(p)\rangle =
\left( p+p^\prime \right)^\mu f_+ + \left( p-p^\prime\right)^\mu f_- \ .
\end{equation}
The choice $a = (q_3, 0, 0, v \cdot q)$ in Eq.~(\ref{bnds}) leads to
the sum rule
\Eqa 
\frac12 (1 &-& \Lam/M_B)^2 + X \frac{\alpha_s}{\pi}
      + {\lambda_1 \over M_B} (\frac1{q_3} 
      -\frac5{6 M_B} ) + {\lambda_2 \over M_B} (\frac1{q_3}
      -\frac1{2 M_B} ) 
\nonumber \\
   &\ge& \  { f_+^2 q_3^2 \over E_\pi M_B}
   \quad \ge \quad  \frac1{E_1 -E_H} \Biggl\{ 
 \frac12 (1 - \Lam/M_B)^2 (E_1 -q_3 -\Lam) + Y \frac{\alpha_s}{\pi} \Delta
\nonumber \\
&+& \frac{\lambda_1}{6 M_B^2} 
  \biggl[5 q_3 -5 E_1 -7 M_B + {M_B \over q_3} (6 E_1 + M_B) \biggr]
\nonumber \\
 &+& \frac{\lambda_2}{4 M_B^2} \biggr[ 2 (q_3 -E_1 +M_B) +
      {M_B \over q_3} (4 E_1 -M_B) \biggr] \Biggr\} \ \  ,
\Endla{fpbd}
where $X$ and $Y$ are both functions of $\Delta$, $q_3$, and
$\Lam$. Their analytic form is rather lengthy and will be presented
in the longer version of this article. We choose $\Delta = 1.5\ \gev$
and $\alpha_s = 0.32$ for all plots in this section.

\begin{figure}
\centerline{
\hfill
\epsfxsize=0.5\textwidth
\epsfbox{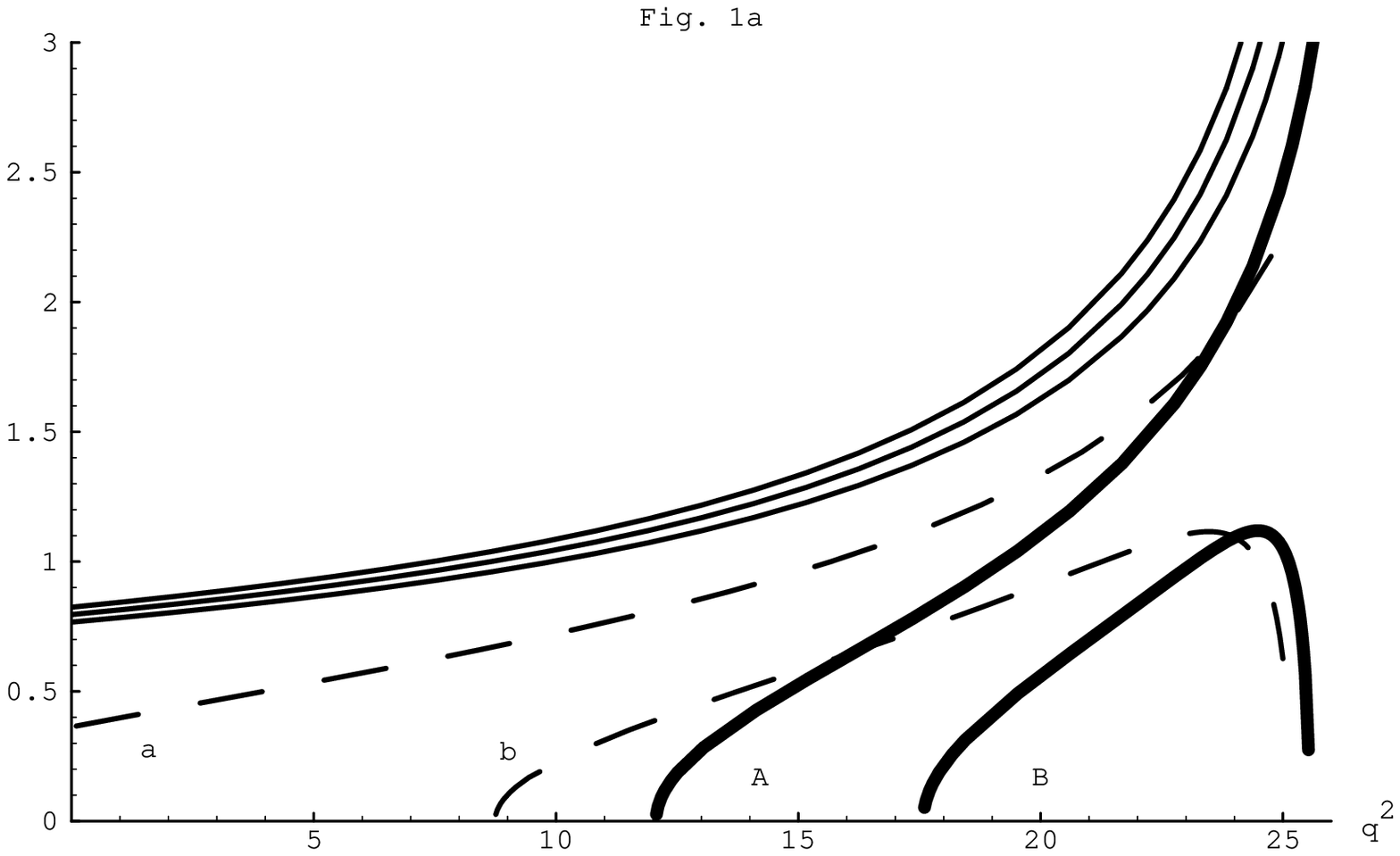}
\hfill
\epsfxsize=0.5\textwidth
\epsfbox{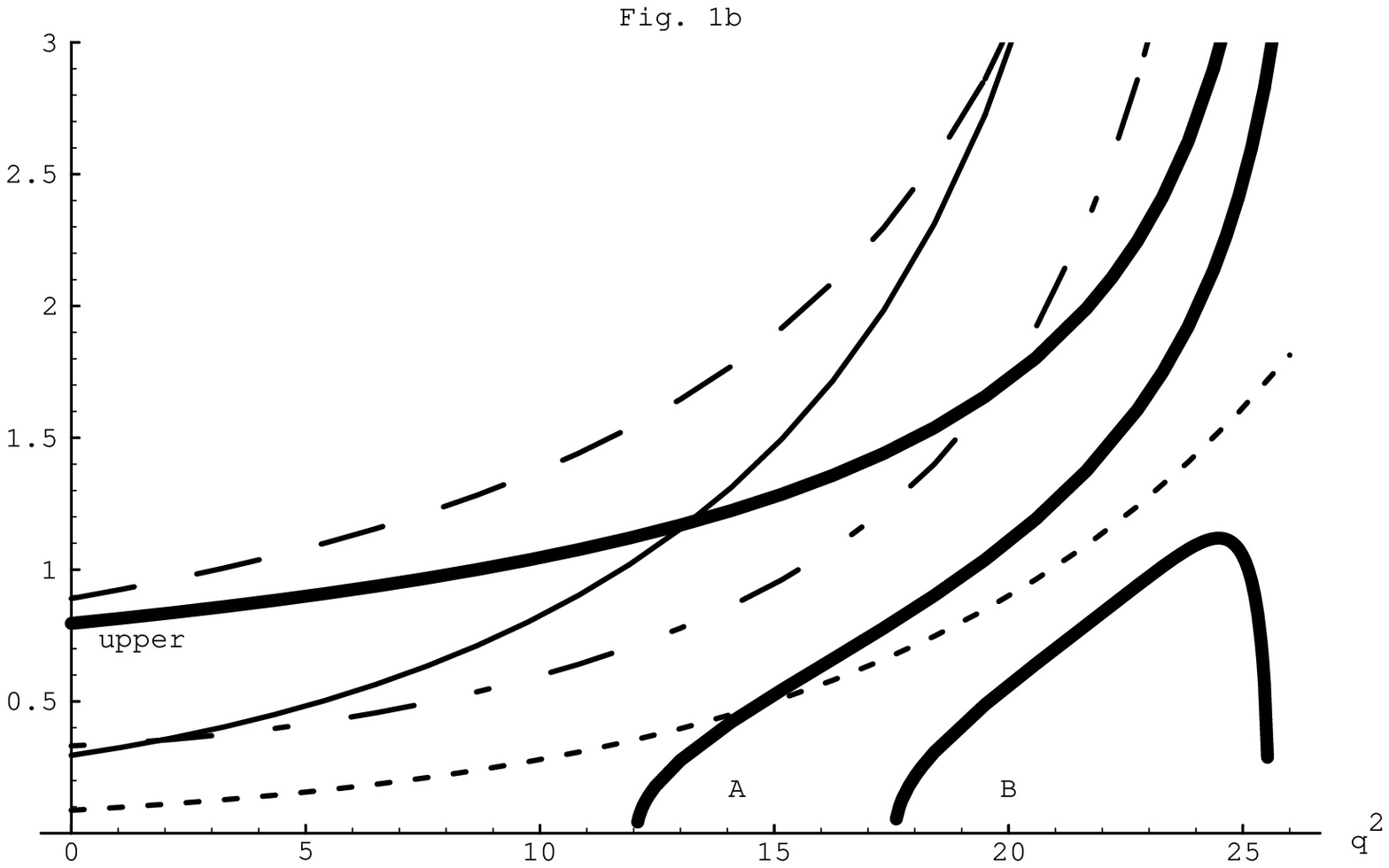}
\hfill}
\caption{\it Upper and lower bounds on the $B \to \pi$ form
factor $f_+$ as a function of momentum transfer $q^2$, in $\gev^2$. 
Fig. 1a : The three thin solid lines correspond to the upper bound for 
data sets A, B, and C as described in the text. The lower bounds
for data sets A and B, respectively, are shown at both tree (dashed a,b) 
and order $\alpha_s$ (thick solid A,B).
Fig. 1b~: Thick solid lines redisplay the upper (labeled ``upper'') and
lower (labeled ``B'') bounds for set B and the lower bound for 
set A (labeled ``A''). The thin lines are model predictions as outlined 
in the text.
}
\label{fig1}
\end{figure}

Figure $1a$ shows the upper and lower bounds on the
form factor $f_+$ as a function of $q^2$, and demonstrates
the importance of various perturbative and nonperturbative 
contributions. The entire kinematic
range has been shown, although the bounds can only be trusted for  
$q^2\ \lesim\ 18\ \gev^2$. The three thin solid
lines correspond to the upper bound with $\Lam$ and $\lambda_1$ 
given by data sets A, B, and C, with
set A being the most constraining, and set C the least. It is evident
the upper bound is rather insensitive to nonperturbative corrections.
Neither is it particularly sensitive to perturbative corrections,
which alter the tree level result (not plotted) by no more than 
$\pm 15 \%$ over the entire $q^2$ range. The tree level lower bounds 
for sets A and B are given by the dashed lines (a) and (b), respectively.
As expected, the one-loop contributions are comparable in size to
the tree level term: they move the dashed curves (a) and (b) into  
the thick solid curves (A) and (B), respectively.  The lower bound for
data set C is not shown because it is unphysical once perturbative
corrections are included (that is, it becomes negative). 

To assess the convergence of the perturbative and nonperturbative
expansions, we look at the upper and lower bounds at
$q^2 = 17.4\ \gev^2$, corresponding to $q_3 = 1\ \gev$.  For data set
A with $\Delta = 1.5\ \gev$ and $ \alpha_s = 0.32$, the perturbative
corrections in Eq.~(\ref{fpbd}) take the values
$X =-0.26$ and $Y = - 0.44$. Numerically, the upper bound is
\Eqa
 { f_+^2 q_3^2 \over E_\pi M_B}\  \le\
  0.500 -  0.055\ \ &+& \  \ 0.008\ -\ \ 0.021\ \  -\ \ 0.006
\nonumber \\
  \CO(1) +  \CO(\Lam) &+&  \CO(\lqcd^2) + \CO(\alpha_s)
       + \CO(\alpha_s \Lam) \ \ ,
\Enda
where the lower line indicates the order in the double
expansion in $\alpha_s$ and $\lqcd$.
Both expansions converge quite well at this value
of $q_3$. For the lower bound, we note that quantities like
$E_1 - q_3 - \Lam$, which represent a mismatch between
perturbative and hadronic endpoints, are $\CO(\lqcd)$. We
evaluate such terms exactly, and for this reason do not show the 
expansion in $\Lam /q_3$, only $\lambda_1/q_3$ and $\lambda_2/q_3$. 
The resulting lower bound is
\Eqa
{ f_+^2 q_3^2 \over E_\pi M_B}\  \ge\
    0.23\qquad &-& \qquad 0.01 \qquad - \qquad 0.12 
\nonumber \\
  \CO({\Lam \over (E_1-E_H)}) &+& 
\CO({\lqcd^2\over (E_1-E_H)}) +  \CO({\alpha_s\over (E_1-E_H)})\ .
\Enda
The nonperturbative corrections appear to be under control. To gauge
the stability of the lower bound with respect to perturbative corrections,
it would be very useful to have an estimate of the two loop contribution (e.g., 
the $\alpha_s^2 \beta_0$ term). In the meantime, it is encouraging 
to note that increasing the one-loop contribution by $25\%$ strongly 
modifies the lower bound (A) only for $q^2\ \lesim\ 15\ \gev^2$, by shifting
the value of $q^2$ where it drops to zero from $q^2 \approx 12\ \gev^2$ 
to $q^2 \approx 14\ \gev^2$ (curve B is modified even less).

The above analysis suggests that both the upper and lower bounds on $f_+$
are reliable for $q^2\ \lesim\ 18\ \gev^2$, once the nonperturbative
parameters $\Lam, \lambda_1$ are given. Are they useful as well? 
In Fig.~$1b$ we again plot the two lower bounds (A) and (B) (thick solid
lines), as well as the upper bound for the central data set B 
(thick solid line labeled ``upper''). Superimposed on these bounds are four
models representing the spread available in the literature. Three
of the models include the contribution of the $B^*$ resonance by using 
heavy meson chiral perturbation theory\cite{hhcpt}, and therefore depend 
on the $B^*$-$B$-$\pi$ coupling $g$ and the $B$ meson decay constant $f_B$
as input parameters.  We adopt values used in the original papers. 
The topmost, dashed line is the prediction of Casalbuoni et al.\cite{Cas}
with $g = 0.61$ and $f_B = .2\ \gev$. With this choice of input parameters,
it is clearly ruled out by our upper bound over the entire kinematic region 
(and the corresponding value of $V_{ub}$ eliminated). The thin solid line 
is the prediction of the Light-Front quark
model by Cheung et al.\cite{CHZ}, with $g = .75$ and $f_B = .187\ \gev$.
It violates the upper bound for $q^2 > 13\ \gev^2$. The dot-dashed curve,
corresponding to the ``constrained dispersive model'' of Burdman 
and Kambor\cite{GK} with $g= 0.5$ and $f_B = .15\ \gev$, is consistent 
with both upper and lower bounds over the kinematic range where the bounds 
are trustworthy. Finally, the dotted curve is the prediction of the
ISGW nonrelativistic quark model\cite{ISGW}. It falls below the 
lower bound (A) at $q^2 = 15\ \gev^2$, so it is inconsistent with the
values $\Lam = .28\ \gev, \lambda_1 = -0.09\ \gev^2$. It has been
pointed out\cite{IWbstar}\ that the ISGW model fails to include
the contribution from the $B^*$ pole at large $q^2$; if modified
to include this, it may well rise above our lower bounds.
The models plotted in Fig.~$1b$ are merely a representative range --
there are many other models and calculations. For example, the
prediction of Bagan \etal\cite{BBB}\ is similar to that of
Burdman and Kambor for $q^2\ \lesim\ 18\ \gev^2$, so it is also consistent
with the bounds.  Lattice simulations\cite{latt} are consistent as well.

While the upper bounds are fairly insensitive to $\Delta$,
the lower bounds can be somewhat improved by smearing over
a smaller range of excitation energies. Taking $\Delta \to 1\ \gev$
lowers the value of $q^2$ where the lower bound vanishes by
$1$ to $2\ \gev^2$. Conversely,
increasing $\Delta \to 2\ \gev$ weakens the lower bound, but only
about half as severely as when $\Delta$ is decreased. We retain 
$\Delta = 1.5\ \gev$ as a reasonable compromise between
maximizing utility and reliability. 

We turn now to the $\bar B \to \rho l \bar \nu$ form factors $a_\pm$ 
defined by
\Eq
\langle \rho(p^\prime)\mid \bar u \gamma^\mu \gamma_5 b \mid\bar B(p)\rangle
= f \epsilon^{*\mu} +\left[
\left(p+p^\prime\right)^\mu a_+ + \left(p-p^\prime\right)^\mu a_-
\right] p\cdot \epsilon^*.
\End
Using $a = (E_\rho, 0,0,-q_3)$ in Eq.~(\ref{masterSR}) 
gives the upper bound
\Eqa 
\frac12 (E_\rho -q_3)^2 &+& \lambda_1 \biggl[ {q_3 \over 3 M_B}
    -{E_\rho \over 3 q_3} - {E_\rho^2 \over 3 q_3 M_B} \biggr]
\nonumber \\
&+& \lambda_2 \biggl[ {q_3 \over M_B } -{E_\rho \over 2 q_3}
       - {E_\rho^2 \over M_B q_3} \biggr] +
\CO(\alpha_s q_3^2) 
\nonumber \\
&\ge& 
{M_B q_3^2 M_\rho^2 \over E_\rho} \Biggl[
a_+ + \frac12 (a_+ + a_-)\biggl({M_B E_\rho \over M_\rho^2}-1 \biggr)
               \Biggr]^2 \ ,
\Endla{apbd}
which is interesting because the leading order term vanishes 
as $q_3$ becomes large. As pointed out in \cite{BR1}, this
implies the $\alpha_s$ contribution dominates at small $q^2$.

It is worth noting that 
Eq.~(\ref{apbd}) implies nontrivial scaling relations for the
form factors $a_\pm$. In the limit that $\lqcd << q_3, M_B$, they 
must fall off at least as fast as 
\Eqa
 a_+ &\sim& {1 \over ( M_B q_3)^{1/2}  } +  { [ \alpha_s(q_3)\ q_3]^{1/2}
      \over  M_B^{1/2} \lqcd }
\nonumber\\
 (a_+ + a_- ) &\sim& {\lqcd^2 \over (M_B q_3)^{3/2} } +
       {  \alpha_s(q_3)^{1/2} \lqcd \over M_B^{3/2} q_3^{1/2} }\ \ .
\Enda
The scaling with $M_B$
follows from considering the heavy $b$-mass limit, which leads 
to\footnote{This relation was used in
\cite{BR1} to simplify Eq.~(\ref{apbd}). This is only valid 
for $b \to c$ decays, since otherwise the smallness of  
$(a_+ + a_-)$ is compensated by $M_B/M_\rho$. }
$a_- = - a_+ [1 + \CO(1/m_b)]$. That $a_+$, modulo perturbative contributions,
falls off with $q_3$ is no surprise, since even a simple pole 
goes like $ 1/q_3$. However, the $q_3$ behavior of $(a_+ + a_-)$
is, to our knowledge, a novel result.
At $q^2=0$, $q_3 \sim M_B$ and the $b$-mass scaling relation need not hold.
Instead, the full $q_3$ scaling relations take over. Practically speaking, 
the large factor $ (M_B E_\rho^{max} - M_\rho^2)/( 2 M_\rho^2) \approx 12$
in Eq.~(\ref{apbd}) leads to very tight constraints on the combination 
$a_+ + a_-$, independent of the validity of the $b$-mass scaling relations.

\begin{figure}
\centerline{
\hfill
\epsfxsize=0.6\textwidth
\epsfbox{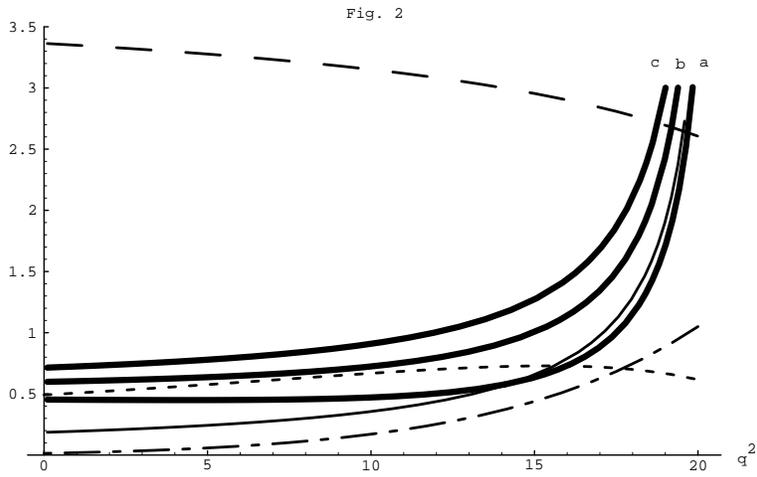}
\hfill}
\caption{\it Upper bounds on the \btorhop\ form factor $A_+$ for data 
sets A, B, C are given by the thick solid curves (a), (b), (c), respectively.
The tree-level upper bound for set B is given by the thin solid curve.
The dashed, dotted, and dot-dashed curves are the model predictions of
WSB, MN, and ISGW, respectively, as 
described in the text.
}
\label{fig2}
\end{figure}

The upper bounds on the magnitude of the dimensionless form factor combination
\Eq
 A_+ \equiv (M_B+M_\rho) \Biggl[ a_+ 
+ \frac12 (a_+ + a_-)\biggl({M_B E_\rho \over M_\rho^2}-1 \biggr) \Biggr]
\Endl{Apdef}
are plotted in thick solid lines
in Fig.~2.  The curves (a), (b), and (c) correspond to data sets A, B, and C,
respectively. The parameter $\lambda_1$ plays an important role in this case
because the leading order $(E_\rho -q_3)^2 $ term suffers from a severe
cancellation. In fact, for $q_3 >> M_\rho$, this term is formally
$\CO( M_\rho^4/q_3^2)$, which is subleading to the contributions from
$\lambda_1$ and $\lambda_2$ (note that $\Lam$ does not enter the upper
bound at tree level). The thin solid line is the tree level upper bound
for $\lambda_1 = -.19\ \gev^2$ (data set B). As expected, the perturbative
corrections dominate at small $q^2$. Even so, changing the one-loop contribution
by $25\%$ alters the upper bounds by no more than $\approx 10\%$ (the square
root helps), so they are reasonably stable against perturbative corrections. 
Again, a two-loop calculation or estimate would be quite welcome here.  

Also plotted in Fig.~2 are three predictions for $A_+$
taken from the literature. The dashed line is the 
relativistic quark model of WSB\cite{WSB}, which exceeds the upper bound
by roughly a factor of five. The reason it violates the bound so spectacularly
is that it does not obey the scaling law $a_- = - a_+$.
Indeed, in order to comply with the bounds, the scaling violation at
$q^2=0$ would have to be reduced by a factor of eight,
$ (a_+ + a_-)/a_+ \lesim\  \frac18 $.
The dotted curve is the prediction of another relativistic quark model
due to Melikhov and Nikitin\cite{MN}, using their ``Set 1'' parameters.
It is consistent with the upper bounds for $\lambda_1 \ge -0.19\ \gev^2$
(their other three sets are consistent for $\lambda_1 \ge -0.09 \ \gev^2$).
The nonrelativistic ISGW quark model\cite{ISGW,IS}\ obeys the 
bounds regardless of the data set. We note that at $q^2=0$,
the model of Cheng \etal\cite{CCH} gives $A_+ = 0.12$, which is
consistent with our bounds. 

Figure 2 illustrates the utility of the inclusive sum rules for constraining,
or even eliminating, models. The sum rules can also yield bounds on the
nonperturbative parameters $\Lam$ and $\lambda_1$\cite{BSUV2,BSUV,GKLW,DL}. 
For example,
the condition $|A_+|^2 \ge 0$ in Eq.~(\ref{apbd}) gives a lower bound
on $\lambda_1$ comparable to \cite{DL}, if we use $\Delta = 1.5\ \gev$. 
The lower bound on $A_+$ also yields a restrictive sum rule, but higher
order nonperturbative terms become important (this is why we have not plotted
the lower bound).  The role of heavy-to-light sum rules in constraining
nonperturbative parameters will be explored in greater detail in the long
version of this article.

\section{Conclusions}

We have presented one-loop improved upper and lower bounds on the
\btopi\ form factor $f_+$, as well as an upper bound on the
combination of \btorho\ form factors $A_+$ defined in Eq.~(\ref{Apdef}).  
We used the upper bounds on
$f_+$, which are exceptionally stable with respect to perturbative and 
nonperturbative corrections, to rule out certain input parameters in
two models.  A third model may be consistent with the lower bounds on
$f_+$, depending on the values of the measurable quantities $\Lam$ and 
$\lambda_1$.  The lower bounds change little if the one-loop
corrections are increased by $25\%$, but an estimation of the two-loop
correction is needed to be certain they are reliable. 

The upper bounds on $A_+$ are moderately stable against further perturbative
corrections, but an estimate of two-loop effects would again be most welcome.
The bounds are particularly interesting because they strongly constrain the 
allowed size of the scale violating (in $m_b$ and $q_3$) quantity $a_+ + a_-$.
Indeed, the model of Wirbel, Stech, and Bauer exceeds the upper bound
by a factor of five. A model by Melikhov and Nikitin obeys the bounds,
but depending on the physical value of $\lambda_1$, their input parameters 
may be constrained. The ISGW model, by contrast, obeys
the bounds rather handily.

The bounds given here demonstrate how inclusive heavy-to-light sum
rules can constrain model parameters, or in some cases even invalidate 
their predictions. They should be useful not only in discriminating between
models, but in constructing, constraining, and fine-tuning them.
A variety of bounds on various combinations
of form factors can be made. The inclusion of $\alpha_s$ perturbative
corrections now allows us to consider, with some reliability, lower bounds
on form factors, upper bounds on heavy quark violating form factor 
combinations, and upper bounds on nonperturbative condensates.  The
results presented here will be extended in a forthcoming publication\cite{BR2},
including analytic expressions for the perturbative corrections.
Improvements to existing sum rules may be possible by either further
restricting the spin-parity of contributing intermediate states, or
by using data to include the contributions of higher states to
the hadronic sum. They should be of use as well in extracting
model-independent limits on the CKM element $V_{ub}$.

\vskip.2in {\centering\large\bf Acknowledgments}

We thank Adam Leibovich for discussions and 
the particle theory group at UCSD for
their hospitality.
This work is supported in part by the Department of Energy under contract 
DOE-ER-40682-137.


\newpage
\dspace{1.8}
\def\np#1{{\it Nucl. Phys.\ }{\bf #1}}
\def\pl#1{{\it Phys. Lett.\ }{\bf #1}}
\def\pr#1{{\it Phys. Rev.\ }{\bf #1}}

\end{document}